\documentclass[11pt,twoside]{article}

\usepackage{asp2014}

\aspSuppressVolSlug
\resetcounters

\begin{document}

\title{Implementing Ideas for Improving Software Citation and Credit}

\author{Peter Teuben,$^1$ Alice Allen,$^{2,1}$ G. Bruce Berriman,$^3$ Kimberly DuPrie,$^{4,2}$ Jessica Mink,$^5$ Thomas Robitaille,$^6$ Keith Shortridge,$^7$ Mark Taylor,$^8$ Rein Warmels$^9$}
\affil{$^1$Astronomy Department, University of Maryland, College Park, MD, US;  \email{teuben@astro.umd.edu}}
\affil{$^2$Astrophysics Source Code Library, College Park, MD, US}
\affil{$^3$Infrared Processing and Analysis Center, California Institute of Technology, Pasadena, CA, US}
\affil{$^4$Space Telescope Science Institute, Baltimore, MD, US}
\affil{$^5$Smithsonian Astronomical Observatory, Cambridge, MA, US}
\affil{$^6$Freelance, Leeds, UK}
\affil{$^7$Knave and Varlet, McMahons Point, NSW, Australia}
\affil{$^8$H.~H.~Wills Physics Laboratory, University of Bristol, U.K.}
\affil{$^9$European Southern Observatory, Garching, Germany}

\paperauthor{Peter Teuben}{teuben@astro.umd.edu}{orcid.org/0000-0003-1774-3436}{University of Maryland}{Astronomy Department}{College Park}{MD}{}{US}
\paperauthor{Alice Allen}{aallen@ascl.net}{}{Astrophysics Source Code Library}{}{College Park}{MD}{}{US}
\paperauthor{G. Bruce Berriman}{gbb@ipac.caltech.edu}{orcid.org/0000-0001-8388-534X}{California Institute of Technology}{IPAC}{Pasadena}{CA}{}{US}
\paperauthor{Kimberly DuPrie}{kduprie@stsci.edu}{}{Space Telescope Science Institute/ASCL}{}{Baltimore}{MD}{}{US}
\paperauthor{Jessica Mink}{jmink@cfa.harvard.edu}{orcid.org/0000-0003-3594-1823}{Smithsonian Astronomical Observatory}{}{Cambridge}{MA}{}{US}
\paperauthor{Thomas Robitaille}{thomas.p.robitaille@gmail.com}{orcid.org/0000-0002-8642-1329}{Freelance}{}{Leeds}{}{}{UK}
\paperauthor{Keith Shortridge}{keithshortridge@gmail.com}{}{Knave and Varlet}{}{McMahons Point}{NSW}{}{Australia}
\paperauthor{M.~B.~Taylor}{m.b.taylor@bristol.ac.uk}{0000-0002-4209-1479}{University of Bristol}{School of Physics}{Bristol}{Bristol}{BS8 1TL}{U.K.}
\paperauthor{Rein Warmels}{rwarmels@eso.org}{0000-0002-9814-0305}{European Southern Observatory}{}{Garching}{}{}{Germany}

\begin{abstract}
Improving software citation and credit continues to be a topic of interest across and within many disciplines, with numerous efforts underway. In this Birds of a Feather (BoF) session, we started with a list of actionable ideas from last year's BoF and other similar efforts and worked alone or in small groups to begin implementing them. Work was captured in a common Google document; the session organizers will disseminate or otherwise put this information to use in or for the community in collaboration with those who contributed. 
\end{abstract}

\section{Introduction}
Though software is increasingly used in research, to the point that all astronomers in an informal survey reported using computational methods in their research \citep{2015arXiv150703989M}, its development and maintenance is not well-rewarded in most disciplines, including astronomy. Several efforts are underway to improve this. Last year, the Center for Open Science published Transparency and Openness Promotion (TOP) Guidelines\footnote{\url{http://centerforopenscience.org/top/}} and earlier this year, the Force11 Software Citation Working Group\footnote{\url{https://www.force11.org/group/software-citation-working-group}} released its Software Citation Principles \citep{softwarecitationprinciples}. The Workshop on Sustainable Software for Science: Practice and Experiences (WSSSPE)\footnote{\url{http://wssspe.researchcomputing.org.uk/}} continues its efforts on software publication, citation, credit, and sustainability \citep{wssspe4}, and a Dagstuhl Perspectives Workshop on Engineering Academic Software\footnote{\url{http://www.dagstuhl.de/de/programm/kalender/semhp/?semnr=16252}} will soon release a manifesto addressing many issues in this area. 

This Birds of a Feather (BoF) session was the latest in a series of ASCL-organized activities at ADASS to find ways to increase the sharing and reuse of, and credit and citation for, software. Starting with a panel discussion at ADASS XXII titled \emph{Bring out your codes! Bring out your codes! (Increasing Software Visibility and Re-use)} \citep{2013ASPC..475..383A}, the ASCL subsequently presented on how to use the ASCL to cite codes \citep{2014ASPC..485..473D} and held a session to generate ideas on how to encourage software sharing \citep{2014ASPC..485....3T} at ADASS XXIII, and at ADASS XXV, held a BoF specifically focused on ways to improve software citation and credit \citep{2015arXiv151207919A}. This session sought to implement some of the actionable ideas that arose from that BoF session and the other efforts mentioned above.

\section{Opening presentation}
Peter Teuben opened the session with remarks about last year's BoF and shared examples of progress made in the past year, such as ADS adding \emph{software} as a document type, thus making it easier to find software in its holdings (a specific request made last year), and Web of Science indexing software records. His brief presentation continued with the action plan for the session and then a list of ideas generated at the previous year's BoF session and that have arisen in other similar discussions in cross-disciplinary meetings such as WSSSPE and the draft manifesto resulting from the Engineering Academic Software workshop held in June 2016. He outlined the schedule for working on these ideas and provided a link for a common Google document to which all could contribute that already contained a list of the shared actionable ideas on which to work. 

\section{Breakout groups}
Participants formed small groups or worked alone to act on the ideas. Keith Shortridge, Bruce Berriman, and Jessica Mink wrote about their experiences in releasing software. Berriman wrote in part on worrying needlessly about thinking ``My code probably has bugs in it'' and his rejoinder, \emph{No, it does have bugs in it}, which had another participant almost immediately commenting in the document, ``This is awesome and very true! Well said!''; obviously, writing about releasing code can have an impact! The texts provided by Shortridge, Berriman, and Mink will be disseminated either as posts on blogs such as the ASCL, Astronomy Computing Today, and AstroBetter, or gathered together into a paper for a journal such as \emph{Journal of Open Research Software}. 

Renato Callado Borges and Greg Sleap provided guidance on the types of software contributions that add value to science. They said that guidance should indicate how to ensure authorship information is properly created or kept, and how to create opportunities to incentivize citation and credit. Among the types of contributions, they listed libraries in general: ``usable classes or functions to solve a particular computing problem. The real value to science is: reusability, confidence/reliability and support (from the community).'' Among other contributions are Docker containers and VM images and operating system distributions such as DebianAstro,\footnote{\url{https://wiki.debian.org/DebianAstro}} which is maintained by Ole Streicher.

Alberto Accomazzi, Nuria Lorente, and Kai Polsterer listed ways one can publish and take credit for software; among them are to
\begin{itemize}                              
\item put code on a public coding repository such as Bitbucket, Github, SourceForge or other resource with a machine-readable codemeta.json CITATION file specifying how use of the code should be acknowledged      
                                   
\item make a release and push the code to a preservation system such as figshare or Zenodo to archive the software and mint a DOI

\item write a conference proceeding paper for ADASS describing the software and post it on arXiv for wider dissemination 

        
\item write a full paper describing the software in an appropriate journal such as \emph{Astronomy \& Computing}, \emph{A\&A} or in AAS journals.
\end{itemize}

Another option is to register the code with the ASCL, which is citable and indexed by ADS (which tracks these citations, something not currently possible if using DOIs for citation); the ASCL is also indexed by Web of Science.

Steven Crawford, Peter Teuben, and possibly others pulled together a list of organization web pages about software created at the institutions to highlight and recognize scientific software contributions, including: 
\begin{itemize} 

\item University of Maryland (\url{https://www.astro.umd.edu/twiki/bin/view/AstroUMD/LocalCodes})

\item Southern African Large Telescope (\url{https://github.com/saltastro/})

\item South African Astronomical Observatory (\url{https://bitbucket.org/saao/profile/repositories}).

\item Space Telescope Science Institute (\url{https://github.com/spacetelescope})
\end{itemize}

Maurizio Tomasi added a suggestion for gathering licensing information and suggested the ASCL require submitters to include how their code is licensed. As a result, several links to licensing information were posted to the document.\footnote{\url{http://ascl.net/wordpress/2015/01/05/software-licensing-resources/}}$^,$\footnote{\url{http://ascl.net/wordpress/2015/01/19/licensing-astrophysics-codes-session-at-aas-225/}}$^,$\footnote{\url{http://www.astrobetter.com/blog/2014/03/10/the-whys-and-hows-of-licensing-scientific-code/}}

Thomas Robitaille, Ole Streicher, Tim Jenness, Kimberly DuPrie, and Alice Allen discussed what the \emph{Preferred citation field} of the ASCL should contain; opinions varied on how much text was reasonable to include and whether acknowledgement (rather than citation) information should appear. In addition, numerous people listed about a dozen preferred citations to be added to the ASCL, either for their own packages or others' software, and other participants used the ASCL's \emph{Suggest a change or addition} link for several software packages to provide preferred citation information.

Teuben had asked people to work for about 30 minutes; in monitoring contributions to the Google doc, he saw work was still being done at the 30-minute mark so let work continue until 15 minutes before the end of the session. With much of the work captured in the Google doc, he did not have participants report on their work and asked instead for other feedback. After a brief discussion on licensing and a few other comments, he closed the session.

\section{Actions taken or planned}
The next phase of this BoF came after the session: putting this information to use or disseminating it. For example, the ASCL has incorporated all of the preferred citations provided and information on institutional websites that list astronomy software is being shared. We will work with participants to determine how best to use other information, continue to take action where we can on the remaining items, and encourage others to do the same. 

\section{Conclusion}
Participants chose ideas for action and worked on them in a common open document. From this input, the organizers of the session drew up action items to get the work done at the BoF used or otherwise implemented or disseminated.

\acknowledgements The ASCL is grateful for financial support provided by the Heidelberg Institute for Theoretical Studies (HITS) and for facilities and services support from Michigan Technological University, the Astronomy Department at the University of Maryland, and the University of Maryland Libraries.

\bibliography{B3}  

\begin{thebibliography}{}
\expandafter\ifx\csname natexlab\endcsname\relax\def\natexlab#1{#1}\fi
\expandafter\ifx\csname url\endcsname\relax
  \def\url#1{\texttt{#1}}\fi
\expandafter\ifx\csname urlprefix\endcsname\relax\def\urlprefix{URL }\fi
\providecommand{\eprint}[2][]{\url{#2}}

\bibitem[{Allen et~al.(2013)}]{2013ASPC..475..383A}
Allen, A., et~al. 2013, in ADASS XXII, edited by D.~N. {Friedel}, vol. 475 of
  ASP Conf. Ser., 383. \eprint{1212.1915}

\bibitem[{Allen et~al.(2015)}]{2015arXiv151207919A}
--- 2015, ArXiv e-prints. \eprint{1512.07919}

\bibitem[{Allen et~al.(2016)}]{wssspe4}
Allen, G., et~al. (eds.) 2016, Proceedings of the Fourth Workshop on
  Sustainable Software for Science: Practice and Experiences (WSSSPE4), no.
  1686 in CEUR Workshop Proceedings (Aachen).
  \urlprefix\url{http://ceur-ws.org/Vol-1686/}

\bibitem[{DuPrie et~al.(2014)}]{2014ASPC..485..473D}
DuPrie, K., et~al. 2014, in ADASS XXIII, edited by N.~{Manset}, \&
  P.~{Forshay}, vol. 485 of ASP Conf. Ser., 473

\bibitem[{{Momcheva} \& {Tollerud}(2015)}]{2015arXiv150703989M}
{Momcheva}, I., \& {Tollerud}, E. 2015, ArXiv e-prints. \eprint{1507.03989}

\bibitem[{{Smith} et~al.(2016){Smith}, {Katz}, {Niemeyer}, \& {FORCE11 Software
  Citation Working Group}}]{softwarecitationprinciples}
{Smith}, A.~M., {Katz}, D.~S., {Niemeyer}, K.~D., \& {FORCE11 Software Citation
  Working Group} 2016, PeerJ Computer Science, 2:e86

\bibitem[{Teuben et~al.(2014)}]{2014ASPC..485....3T}
Teuben, P., et~al. 2014, in ADASS XXIII, edited by N.~{Manset}, \&
  P.~{Forshay}, vol. 485 of ASP Conf. Ser., 3. \eprint{1312.7352}

\end{thebibliography}

\end{document}